\documentstyle[prb,aps,twocolumn,epsfig,amsmath]{revtex}

\begin{document}
\title{Newton's law for Bloch electrons, Klein factors and
  deviations from canonical commutation relations }

\author {K.Sch\"onhammer}
\address{Institut f\"ur Theoretische Physik, Universit\"at
  G\"ottingen, Bunsenstr. 9, D-37073 G\"ottingen, Germany}

\date{\today}
\maketitle

\begin{abstract}
The acceleration theorem for Bloch electrons in a homogenous external
field is usually presented using quasiclassical arguments. In 
quantum mechanical versions the Heisenberg equations of motion for
an operator  $\hat {\vec k}(t)$ are presented mostly without properly
defining this operator. This leads to the surprising fact that the
generally accepted version of the theorem is incorrect for the most
natural definition of $\hat {\vec k}$. This operator is shown {\it not}
to obey canonical 
commutation relations with the position operator. A similar 
result is shown for the phase operators defined via the {\it Klein factors}
which take care of the change of particle number in the {\it bosonization
of the field
operator} in the description of
interacting fermions in one dimension. The phase operators are
also shown
not to obey canonical commutation relations with the
corresponding particle number operators.  
Implications of this fact are discussed for Tomonaga-Luttinger
type models.   

\end{abstract}
\section{Introduction}
The problem of the electronic motion in a crystal subject to a
homogeneous electric field is treated in virtually all textbooks on
solid state physics \cite{Ki,AM}. Using semiclassical arguments one
arrives at the $\vec k$-space form of Newton's law for Bloch waves
\begin{equation}
\hbar \frac{d \vec k}{dt} = \vec F
\end{equation}
where $\vec F = e \vec E$, with $\vec E$ the electric field, is a
spatially uniform force $\vec F$. In more advanced texts discussions
are presented that this law also holds rigorously as a quantum
mechanical law \cite{QKi,Ca}, when $\vec k(t)$ is interpreted as the
expectation value of an {\it operator} $\hat {\vec k}(t) = e^{iHt}\hat
{\vec k}
e^{-iHt}$ for an arbitrary wave packet $|\psi\rangle$. In this paper we will
point out various subtleties in the definition of the operator $\hat
{\vec k}$, which surprisingly are not sufficiently discussed in the
literature.

It is well-known that there are certain mathematical
problems with the fact that the perturbation $-\vec F \cdot \hat{\vec
  x}$ is an {\it unbounded} operator \cite{Av,Ne}. This has led to a
long discussion about the existence of 
localized eigenstates, so-called
``Wannier-Stark-ladders'' \cite{Wa,Kri}. It has been shown that the
problem is very different if one restricts the Hilbert space to a
{\it finite} number of Bloch bands or works in the {\it full} Hilbert
space \cite{Av}. As the important aspects of the problem show up
already in {\it one-dimensional} systems we restrict our discussion to this
case. A little more mathematical sophistication than usual in solid
state physics is required to appreciate the importance of a proper
definition of the operator $\hat k$. Take for example the single band
case discussed in section III and let $|E_0\rangle$ be a localized
Wannier-Stark eigenstate of the Hamiltonian. Then obviously $\langle E_0|\hat
k(t)|E_0\rangle = \langle E_0|\hat k|E_0\rangle$,
 i.e. the expectation value is {\it time-independent}
 in contradiction to Eq. (1). The solution to this problem
has some similarity to the subtleties of the canonical commutation
relations for a particle on a ring \cite{J,K}.
But there the wave function itself is defined on a compact space,
while this is not the case for the Bloch electron in a homogenous
field where the periodicity enters via the $k$-space classification of
the eigenstates in a periodic potential.
 A similar problem
occurs in a proper definition of the so-called ``Klein-factors'' $U$,
which in the method of ``bosonization'' \cite{Ha,GNT,DS,KS} take proper
care of the change of the number of particles, when the electronic
field operator is expressed in terms of boson operators. The $U$ are
unitary operators and as such can be expressed as $U = e^{iM}$, where
$M$ is a self-adjoint operator. We will show in this paper, that,
given $U$ the operator $M$ remains undetermined to a larger extent
than one could naively expect. But whatever choice one takes, this
operator $M$ can be shown \cite{KS} {\it not} to obey the commutation
relation with the particle number which is widely assumed \cite{Ha}.

In section II we discuss the problem of an electron in a periodic
potential and a homogeneous external field. First introducing the
translation operator $\hat{T}_a$ by one lattice spacing, different
definitions of the operator $\hat k$ are given, which all lead to
$\hat T_a = e^{-i{\hat k}a}$, but different forms of Newton's law and
commutation relations with the position operator. The most obvious
choice for $\hat k$ does {\it not} lead to Newton's law in the form of
Eq. (1) and allows to resolve the contradiction 
for the case of Wannier-Stark states mentioned above.
In section III we discuss the problem, when the Hilbert space
is restricted to a {\it finite} number of bands. The results
of sections II and III can directly be used
  for the discussion of the Klein operators in section
IV. We finally discuss the implications in the rather different
situations where the same mathematical problem shows up in section V.

\section{A Bloch electron in a homogenous field}
In this section we discuss the motion of an electron in a periodic
potential $V(x) = V(x+a)$, where $a$ is the lattice constant, subject
to an additional homogeneous external field $F$
\begin{equation}
\hat H = \frac{\hat p^2}{2m} + V(\hat x) - F\hat x.
\end{equation}
In the attempt to derive the $k$-space form of Newton's law it is
useful to introduce the translation operator \cite{Kr}
which (actively) shifts the state by one lattice spacing to the right
$\langle x|\hat T_a|\psi\rangle = \langle x-a|\psi\rangle$. The commutator 
of $\hat T_a$ with the Hamiltonian
is given by 
\begin{equation}
[\hat T_a, \hat H] = [\hat T_a, -F \hat x] = a F\hat T_a
\end{equation}
Therefore the Heisenberg equation of motion $i\hbar \dot{\hat T}_a(t) =
[\hat T_a,\hat H](t)$
has the solution
\begin{equation}
 \hat T_a(t) = \hat T_a e^{-iaFt/\hbar}.
\end{equation}
As it is well known from elementary quantum mechanics books the most
obvious way to present the translation operator is
$T_a=e^{-ia\hat p/\hbar}$, 
where $\hat p$ is the momentum operator. This is
{\it not} the appropriate choice in the following. We nevertheless show
which contradictions one seems to produce using naive reasoning.
With the translation operator represented in terms of the momentum
operator Eq. (4) reads
\begin{equation}
e^{-ia\hat p(t)/\hbar} = e^{-ia(\hat p + Ft\hat 1)/\hbar}.
\end{equation}
Obviously it is {\it not} allowed to infer $\hat p(t) = \hat p + Ft
\hat 1$
from this equation, as it contradicts for all $V \neq 0$ the
Heisenberg equation of motion
\begin{equation} 
\dot{\hat p}(t) = -\frac {\partial V}{\partial x}(\hat x(t)) + F\hat 1.
\end{equation}
We postpone the discussion of what goes wrong with this argument
 and first discuss the case $V \equiv 0$, i.e.
a system {\it
  without} the periodic potential (``empty lattice''),
where {\it no} contradiction arises. The spectral decomposition
of $\hat k \equiv \hat p/\hbar$ reads
\begin{equation}
\hat k = \int^\infty_{-\infty} |k\rangle k\langle k|dk\;,
\end{equation}
where the $|k\rangle$ are plane wave states $\langle x|k\rangle = e^{ikx}/\sqrt{2\pi}$,
which are normalized as  $\langle k|k'\rangle = \delta(k-k')$. In order to find
$\hat k(t)$ 
we do not proceed the simple way using Eq. (6) but
 calculate $|k(-t)\rangle\langle k(-t)|$,
needed later. The operator
$\hat P_{0,k} \equiv |k\rangle\langle k|$
 commutes with the kinetic energy and its
commutator with $\hat x$ follows using $\langle k|\hat x |\psi\rangle = i
\frac{\partial}{\partial k}\langle k|\psi \rangle$
 as $[\hat P_{0,k}, \hat x] = i
\frac{\partial}{\partial k} \hat P_{0,k}$.  For $V = 0$ this leads to the
equation of motion
\begin{equation}
i\hbar \dot{\hat P}_{0,k}(t) = -iF\frac{\partial}{\partial k}\hat P_{0,k}(t)
\end{equation}
with the solution $\hat P_{0,k}(t) = \hat P_{0,k-Ft/\hbar }(0)$,
i.e. $|k(-t)\rangle\langle k(-t)|= |k-Ft/\hbar \rangle\langle k-Ft/\hbar |$.
 Using Eq. (7) this yields the result expected from Eq. (6)
\begin{equation}
\hat k(t) = \int^\infty_{-\infty}|k-Ft/\hbar\rangle k\langle
k-Ft/\hbar|dk =
 \hat k +   Ft/\hbar {\hat 1}
\;.
\end{equation} 
This shows explicitly that for $V= 0$ just comparing the
exponents in Eq. (5) provides the correct answer. 

We now address the question of how much an operator $\hat{\tilde k}$
can differ from $\hat k$ defined in Eq. (7) and nevertheless yield
$\hat T_a = e^{-i{\hat{\tilde k}}a}$. To answer this we consider an
arbitrary projection operator $\hat P = \hat P^2$. Simply expanding
the exponential function leads to the identity
\begin{equation}
e^{i\alpha \hat{P}} = \hat 1 + \left(e^{i\alpha} - 1\right)\hat P.
\end{equation}
For $\alpha = 2\pi n$, with $n$ integer this simplifies to 
\begin{equation}
e^{2\pi ni\hat P} = \hat 1. 
\end{equation}
If two operators $\hat A$ and $\hat B$ fulfill $e^{i\hat A} = e^{i\hat
  B}$
multiplication with $e^{2\pi ni \hat P}=\hat 1$ yields $e^{i\hat A}
 = e^{i\hat B}e^{2
  \pi in \hat P}$. If $\hat P$ {\it commutes} with $\hat B$, one obtains
$e^{i\hat A} = e^{i(\hat B+2 \pi ni \hat P)}$,
 i.e. $\hat A$ and $\hat B$ can differ by
$2\pi n \hat P$ and still fulfill $e^{i\hat A} = e^{i\hat B}$. If
$\hat B$ is a
self-adjoint operator with the spectral decomposition
\begin{equation}
\hat B = \sum_b |b\rangle b\langle b|
\end{equation}
every {\it restricted} sum (integral) $\hat P = \sum'_b |b\rangle\langle b|$ is a
projector which commutes with $\hat B$. This simple observation is
essential for the problem addressed in this paper.

As a first application we write $k$
in the integrand
 in Eq. (7) in the form $k = \tilde k +
\frac{2\pi}{a}n$, with $n$ integer and $\tilde k \in [-\pi/a, \pi/
a]$. Then we can rewrite this equation in the form
\begin{eqnarray}
\hat k & = &
\sum^\infty_{n=- \infty}\int^{\pi/a}_{-\pi/a}|k + \frac{2\pi}{a}n\rangle(k +
\frac{2\pi}{a}n)\langle  k + \frac{2\pi}{a}n|dk\nonumber\\ 
& = & \sum^\infty_{n=-\infty} \int^{\pi/a}_{-\pi/a}|k +
\frac{2\pi}{a}n\rangle k \langle k + \frac{2\pi}{a}n|dk\nonumber\\
& +  & \sum^\infty_{n = -\infty}\frac{2\pi}{a}n \int^{\pi/a}_{-\pi/a}|k +
\frac{2\pi}{a}n\rangle\langle k 
+ \frac{2\pi}{a}n|dk \nonumber\\
&\equiv& \hat k_b^{(a)}+ \sum^\infty_{n = -\infty}\frac{2\pi n}{a}\hat P^{(n)},
\end{eqnarray}
where now $\hat k_b^{(a)} $ is a {\it bounded} operator, and the projection
operators $\hat P^{(n)}$ commute with $\hat k_b^{(a)}$, i.e. we also have 
$\hat T_a = e^{-ia\hat k_b^{(a)}}$.
In the position representation $\hat k_b^{(a)}$
 is a {\it nonlocal} operator
\begin{eqnarray} 
\lefteqn{
\langle x|\hat k_b^{(a)}|\psi \rangle=
\int_{-\infty}^{\infty}\frac {1}{2\pi}\sum_{n=-\infty}^{\infty}
e^{i\frac {2\pi}{a} n(x-x')}
}\nonumber\\
& &\hspace{20ex}\times \left ( \int_{-\pi /a}^{\pi /a}ke^{ik(x-x')}dk \right )
  \psi(x') dx'\nonumber\\
& & =
\frac {1}{i} \int_{-\infty}^{\infty} \sum_{m=-\infty}^{\infty}
\delta (x-x'-ma)\nonumber\\
& & 
\times \left [ \frac {\cos{ \left [(x-x')\pi/a\right ]}}{(x-x')}
-\frac {a\sin{\left [(x-x')\pi /a \right ]}}{\pi (x-x')^2}\right ]\psi (x')dx'
\nonumber\\
&  & =
\sum_{m=1}^{\infty}\frac {(-1)^m}{iam}\left
  [\psi(x-ma)-\psi(x+ma)\right ],
\end{eqnarray}
\noindent where we have used Eq. (A1). In the following we suppress
the
$a$-dependence of the operator $k_b^{(a)}$. 
In order to discuss $\hat k_b(t)$ it is convenient to introduce the
operators
\begin{equation}
\hat P_k \equiv \sum^\infty_{n = -\infty} |k + \frac{2\pi}{a}n\rangle \langle k +
\frac{2\pi}{a}n|= \hat P_{k + \frac{2\pi}{a}m}\, , 
\end{equation}
where we have indicated that they are periodic in the reciprocal
lattice ($m \in \mathbf Z$) and we have
\begin{equation}
\hat k_b = \int^{\pi/a}_{-\pi/a} \hat P_k kdk\;.
\end{equation}
If for $V=0$ we again use $|k(-t)\rangle\langle k(-t)| = |k-Ft/\hbar
\rangle\langle k-Ft/\hbar|$ we obtain
$\hat P_k(t) = \hat P_{k-Ft/\hbar}$. Because of the periodicity $\hat P_k =
\hat P_{k + \frac{2\pi}{a}m}$ we have $\hat P_k(t) =\hat P_k(t+m T)$ where
$T = 2\pi\hbar/(aF)$, and we can restrict the following discussion to times
$t \in [0,T]$
\begin{eqnarray}
\hat k_b (t)
& = & 
\int^{\pi/a}_{-\pi/a} \hat P_{k-Ft/\hbar }kdk \nonumber\\
& = &
\int^{-\pi/a + Ft/\hbar }_{-\pi/a}\hat P_{k-Ft/\hbar + 2\pi/a}kdk
\nonumber\\
& & \hspace{5ex} + \int^{\pi/a}_{-\pi/a + Ft/\hbar } \hat P_{k-Ft/\hbar }kdk,
\end{eqnarray}
where we have used Eq. (15). Now we can substitute $k' = k-Ft/\hbar  +
2\pi/a$ in the first term and $k' = k - Ft/\hbar $ in the
second. This yields 
\begin{eqnarray}
\hat k_b(t) 
& = & \int^{\pi/a}_{\frac{\pi}{a}-Ft/\hbar }
\left (k' + Ft/\hbar  -
2\pi/a\right )    \hat P_{k'}     dk'\nonumber\\
& & + \int_{-\pi/a}^{\frac{\pi}{a}-Ft/\hbar } \left(k' + Ft/\hbar \right )
    \hat P_{k'}   dk'
\end{eqnarray}
Apart from the term proportional to $2 \pi/a$ we can recombine the
integrals and obtain using $\int^{\pi/a}_{-\pi/a}\hat P_k dk = \hat 1$
\begin{equation}
\hat k_b(t) = \hat k_b + Ft/\hbar \hat 1 - \frac{2
  \pi}{a}\int^{\pi/a}_{\pi/a - Ft/\hbar }\hat P_k dk\, .
\end{equation}
Comparing with Eq. (9) we see that the expressions of
 $\hat k(t)$ and $\hat k_b(t)$ in terms of $\hat k$ and $\hat k_b$
differ by the last term on the rhs of Eq. (19). We will discuss its
implications after we show that in fact Eq. (19) also holds in the presence
of the periodic potential. We denote the 
eigenstates of $\hat H_0 = \hat p^2/2m + V(\hat x)$ by $|k,\alpha \rangle $
\begin{equation}
\hat H_0 |k,\alpha\rangle = \epsilon_{k,\alpha}|k,\alpha\rangle\,,
\end{equation}
where $k$ is the wave vector in the first Brillouin zone and $\alpha$
is the band index. When the states are normalized as
$\langle k,\alpha|k',\alpha'\rangle = \delta_{\alpha,\alpha'}\delta(k-k')$ they
provide a decomposition of the unit operator
\begin{equation}
\sum_{\alpha}\int_{-\pi/a}^{\pi/a}|k,\alpha\rangle\langle k,\alpha|dk = \hat 1\,.
\end{equation}
In the appendix we present the simple proof that the $\hat P_k$ can
also be expressed in terms of the Bloch states
\begin{equation}
\hat P_k = \sum^\infty_{n = -\infty}|k+
\frac{2\pi}{a}n\rangle\langle k + \frac{2\pi}{a}n| = 
\sum_\alpha |k,\alpha\rangle\langle k,\alpha|
\end{equation}
The Heisenberg equation of motion for $\hat P_k(t)$ reads 
\begin{equation}
i\hbar \dot {\hat P}_k(t) = \left[\hat P_k,\hat H_0\right](t)
 - F\left[ \hat P_k,\hat
  x\right](t) 
\end{equation}
If one uses the expression for $\hat P_k$ in terms of the Bloch states
the first term on the rhs is seen to vanish, while the expression
using the plane waves shows that $\left[\hat P_k,\hat x\right] = i
\frac{\partial}{\partial k}\hat P_k$ and Eq. (23) simplifies to
$\hbar 
\dot{\hat P}_k(t) + F \frac{\partial}{\partial k} P_k(t) = 0$. This
leads to the solution
\begin{equation}
\hat P_k(t) = \hat P_{k-Ft/\hbar }(0),
\end{equation}
which expresses the 
well known fact that a state with a single sharp value of $k$
remains a state with a single sharp value of $k$, independent of any
interband transitions \cite{QKi,Ca,Kr}.
The
 further steps for the calculation of $\hat k_b(t)$ 
are identical to
Eqs. (16-19). Therefore Newton's law for Bloch electrons in a
homogeneous field has the solution $\hat k_b(t) = \hat k_b(t + nT)$ and
for $t \in [0,T]$
\begin{equation}
\hat k_b(t) = \hat k_b + Ft/\hbar  \hat 1 - \frac{2\pi}{a}
 \sum_{\alpha} \int^{\pi/a}_{\pi/a-Ft/\hbar 
}|k,\alpha\rangle\langle k,\alpha|dk
\end{equation}
This one of the central results of the present paper. 
The last term on the rhs of Eq. (25) is missing in all presentations
of the ``acceleration law'' known to me.
Its appearance can be traced to the fact that the operators
$\hat P_k$ are {\it periodic } in the reciprocal lattice.
 A similar term has been
discussed previously in connection with a free particle on a ring
\cite{J}.

\noindent If one calculates expectation values with a general wave
packet $|\psi\rangle = \sum_\beta \int^{\pi/a}_{-\pi/a}
 a_\beta (k)|k,\beta \rangle
dk$ the expectation value of $\langle \hat k_b(t)\rangle = \langle  \hat k_b(t + nT)\rangle$ for
$t \in [0,T]$ is given by
\begin{equation}
\langle \hat k_b(t)\rangle = \langle \hat k_b\rangle + Ft/\hbar  - \frac{2\pi}{a} \sum_\beta
\int^{\pi/a}_{\pi/a - Ft/\hbar }|a_\beta(k)|^2 dk,
\end{equation}
with $\langle \hat k_b\rangle = \sum_\beta \int^{\pi/a}_{-\pi/a}|a_\beta (k)|^2
kdk$. For a wave packet
initially strongly peaked around a wave number $k_0$,
the expectation value $\langle \hat k_b(t)\rangle$ closely follows the saw-tooth
curve shown in Fig. 1 except at the ``times of the
Bragg-reflections''. This implies that Eq. (1) holds to a good 
approximation except during these reflections, when $k$ is identified
with $ \langle \hat k_b(t)\rangle $. For an {\it arbitrary} wave packet
this is {\it not}   \cite {CaII}   the case due to the additional term from
differentiating the last term on the rhs of Eq. (26).
This can be clearly seen in Fig. 1.

\begin{figure}[hbt]
\begin{center}
\vspace{-0.0cm}
\leavevmode
\epsfxsize6.5cm
\epsffile{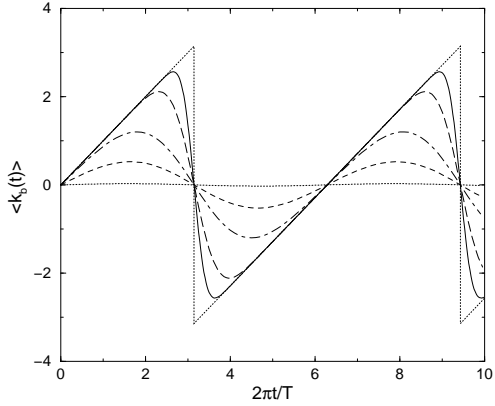}
\caption{ Expectation value of the operator $\hat k_b(t)$
as a function of $2\pi t/T=taF/\hbar$ 
for a normalized Gaussian wave packet
$|\psi \rangle =c_0 \sum_n \exp (-\alpha n^2 )|n\rangle $, where
$|n\rangle$ are the Wannier states in a single band model. The values
for $\alpha$ are 5 (dotted), 2 (dashed), 1 (dashed dotted), 0.2 (long
dashed) and 0.05 (solid). The asymptotic saw tooth curve is also shown 
(dotted). }
\label{kvt}
\end{center}
\vspace{0.0cm}
\end{figure}

\noindent The additional term on the rhs of Eq. (25) allows
to resolve the contradictions mentioned in the introduction. This will
be discussed in the next section. It also leads to an {\it additional}
term in the commutator $[\hat x,\hat k_b]$ compared to $[\hat x,\hat
k] = i \hat 1$. If we compare the Heisenberg equation
of motion $i\dot{\hat
  k_b}(t) = \left[k_b, H\right](t) = -F \left[\hat k_b,\hat
  x\right](t)$ for $t = 0$ with Eq. (25) differentiated with respect
to time we readily obtain
\begin{equation}
\left[\hat x, \hat k_b\right] = i\left (\hat 1 -
\frac  {2\pi}{a} \sum_\alpha
|\frac{\pi}{a},\alpha\rangle\langle  \frac{\pi}{a},\alpha|\right ),
\end{equation}
i.e. $\hat x$ and $\hat k_b$ do {\it not} obey canonical commutation
relations.
If one goes over from  $\hat k_b$ to $\hat {\tilde k}_b$
by adding $2\pi in\hat P/a$, where $\hat P$ is a projection operator 
commuting with  $\hat k_b$, additional terms appear on the rhs
 of Eq. (27).\cite{Beispiel}
 A detailed discussion of the additional term is given in
the next section, where we also present an alternative derivation of
the result.

\section{Description in a restricted Hilbert space}
The description of a Bloch electron in the external field simplifies
considerably when the dynamics of an initial state in the form 
 of a wave packet in a {\it single} band is
considered and interband transitions are neglected \cite{Wa,Kri}. In
this approximation the electron undergoes ``Bloch oscillations''
i.e. ${\langle \hat x (t)\rangle}$ is periodic in time.
 A first step beyond this
simple approximation is a two-band model \cite{FBF} which provides a
good description for the dynamics of electrons in high quality
semiconductor superlattices, with a pair of isolated minibands
\cite{H,Z}. More generally the $N$-band approximation amounts to the
replacement
\begin{equation}
\hat H_0 \to
\sum^N_{i=1}\int^{\pi/a}_{-\pi/a}|k,\alpha_i\rangle\varepsilon_{k,\alpha_{i}}
\langle k,\alpha_i|dk\; ,
\end{equation}
and the restriction of the operator $\hat x$ to the Hilbert space
spanned by the $N$ bands. In the following it is useful also to work
with Wannier states
\begin{equation}
|m,\alpha\rangle = \sqrt{\frac{a}{2\pi}} \int^{\pi/a}_{-\pi/a} e^{-ikam}
|k,\alpha\rangle dk,
\end{equation}
which are localized around the $m$-th lattice site,
 i.e. $\langle x|m,\alpha\rangle
=a_\alpha (x-ma)$. They are orthonormal $\langle m,\alpha|n,\beta\rangle =
\delta_{\alpha \beta} \delta_{mn}$ and therefore can also be used for
a decomposition of the unit operator in the restricted $N$-band
Hilbert space. Elementary calculation yields for the Wannier matrix
elements of the position operator
\begin{eqnarray}
\langle m,\alpha|\hat x|n,\beta\rangle 
& = &
ma \delta_{mn} \delta_{\alpha \beta} + \Omega^{m-n}_{\alpha \beta}
\nonumber\\
\Omega^{m-n}_{\alpha \beta} 
& = &
\int^{\infty}_{-\infty} dx\; a^*_\alpha (x) x\; a_\beta (x+(m-n)a)
\end{eqnarray}
with $(\Omega_{\alpha \beta}^{m-n})^* = \Omega^{n-m}_{\beta \alpha}$,
which implies for the Bloch state matrix elements using (A.1)
\begin{equation}
\langle k,\alpha|\hat x| k',\beta\rangle = \delta_{\alpha \beta} i
\frac{\partial}{\partial k} \delta_{2\pi/a}(k-k') + \delta_{2
  \pi/a}(k-k') \Omega_{\alpha \beta}(k),
\end{equation}
with $\Omega_{\alpha \beta}(k) = \sum_l e^{-ikal} \Omega^l_{\alpha
  \beta} = \left[\Omega_{\beta \alpha}(k)\right]^*$
and      $ \delta_{2\pi/a}(\cdot)$ the $2\pi/a$-periodic delta function defined
in (A1). In the following
we suppress the index ``$2\pi/a$'' of the delta functions as wave
numbers are always restricted to the first Brillouin zone. 

The translation operator $\hat T_a$ when acting on a Wannier
state changes $m$ by one, i.e. $\hat T_a|m,\alpha\rangle =
|m+1,\alpha\rangle$.
 Therefore
the restriction $T_a^{(N)}$ on the $N$-band space has the simple form
\begin{equation}
\hat T_a^{(N)} = \sum^N_{i=1}\sum^\infty_{m = -\infty}
|m + 1, \alpha_i\rangle\langle m,\alpha_i|
\end{equation}
From the inverse relation of Eq. (29)
\begin{equation}
|k,\alpha\rangle = \sqrt{\frac{a}{2\pi}} \sum^\infty_{m = -\infty}
e^{ikam}|m,\alpha\rangle
\end{equation}
it follows that $\hat T_a^{(N)}|k,\alpha\rangle = e^{-ika}|k,\alpha\rangle$ and
therefore $\hat T_a^{(N)}$ can be written
 as $\hat T_a^{(N)} = \exp {(-ia \hat
k_b^{(N)})}$ with
\begin{equation}
\hat k_b^{(N)} = \sum^N_{i=1} \int^{\pi/a}_{-\pi/a}
|k,\alpha_i\rangle\langle k,\alpha_i|kdk \equiv \int^{\pi/a}_{-\pi/a} \hat P_k^{(N)}kdk\;.
\end{equation}
The only difference to section II is therefore the restricted sum over
bands. In the equation of motion for $\hat P_k^N(t)$ one again has to
determine the commutator with the position operator. As now there is no
equivalent to Eq. (22) the argument presented in appendix A cannot be
used. Using Eq. (31) it is straightforward to show that 
\begin{equation}
\left[ \hat x^{(N)}, \hat P_k^{(N)}\right] = i
\frac{\partial}{\partial k} \hat P_k^{(N)}
\end{equation}
by calculating matrix elements in Bloch states of both sides of this
equation. Then all further arguments of section II can be used and one
obtains $\hat k_b^{(N)}(t) = \hat k_b^{(N)}(t + n T)$ and for $t \in
[0,T]$ 
\begin{eqnarray}
\hat k_b^{(N)}(t) 
& = & \hat k_b^{(N)} +   \hat 1^{(N)}   Ft/\hbar \nonumber\\ 
& & - 
\frac{2 \pi}{a} \sum^N_{i=1} \int^{\pi/a}_{\pi/a -
  Ft/\hbar }|k,\alpha_i\rangle\langle k,\alpha_i|dk\;. 
\end{eqnarray}
Again the only difference to section II is the restricted sum over
bands. The main new physics is that for the restricted $N$-band
Hilbertspace {\it localized} eigenstates $|E_n^\mu\rangle =
\left(\hat T_a^{(N)}\right)^n |E^\mu_0\rangle$ of $\hat 
H_0^{(N)} - F\hat x^{(N)}$,
with $n \in \mathbf Z$ and $\mu = 1,...N$ are known to exist
\cite{Av,Ne}. The corresponding energies $E^\mu_n$ form $N$
``Wannier-Stark'' ladders
\begin{equation}
E^\mu_n = E_0^\mu - naF.
\end{equation}
This property is easily shown using $\left(\hat T_a^{(N)}\right)^{-1}
\hat H^{(N)}\hat T_a^{(N)} = \hat H^{(N)} - Fa \hat 1$. Now we can resolve the
contradiction of $\langle E^n_\mu|\hat k_b^{(N)}(t)|E^n_\mu\rangle = \langle E_\mu^n |\hat
k_b^{(N)}|E^n_\mu\rangle$ with Eq. (1) mentioned in the introduction. The
last term on the rhs of Eq. (36) allows to
avoid the inconsistency. If one takes the expectation value of Eq. (36)
with the normalized eigenstate $|E_n^\mu\rangle$ and differentiates with
respect to time one finds that ${d\langle E_n^\mu|\hat
k_b^{(N)}(t)|E_n^\mu\rangle/dt = 0}$ is fulfilled if 
\begin{equation}
\sum^N_{i=1}|\langle k,\alpha_i|E^\mu_n\rangle|^2 = \frac{a}{2\pi}
\end{equation}
holds. In order to see that this in fact is true one has to examine the
time-independent Schr\"odinger equation $\hat H^{(N)}|E^\mu_n\rangle =
E^\mu_n|E_n^\mu\rangle$. Using Eq. (31) it reads in Bloch representation
\begin{eqnarray}
i \frac{\partial}{\partial k}\langle k,\alpha_i | E^\mu_n\rangle 
& = & 
-\left(\frac{E_n^\mu - \varepsilon_{k,\alpha_i}}{F}\right)\langle k,
 \alpha_i |E_n^\mu\rangle \nonumber\\
& &
-\sum^N_{j=1} \Omega_{\alpha_{i}\alpha_{j}}^{(k)} \langle
k,\alpha_j|E^\mu_n\rangle\;. 
\end{eqnarray}
With $k \to t$ this looks like a ``time'' dependent Schr\"odinger
equation with a ``time'' dependent Hermitian $N \times N$ Hamiltonian
matrix. The ``conservation of probability'' immediately shows that the
lhs of Eq. (38) is $k$-independent. Integration over the first Brillouin
zone allows to use the completeness relation. Because of
$\langle E^\mu_n|E_n^\mu\rangle = 1$ this immediately proves Eq. (38). In the
simplest case $N = 1$ Eq. (39) is a single first order linear
differential equation and can be trivially integrated and allows an
analytical discussion of the eigenstates $|E_n\rangle$ in real space
\cite{FBF}.

It is instructive to write the operator $\hat k_b$ in Eq. (34) also in
the Wannier representation. If one performs the simple integration
as in Eq. (14) in 
$\langle m,\alpha|\hat k_b^{(N)}|n,\beta\rangle = \delta_{\alpha \beta}\langle m,\alpha
|\hat k_b^{(N)}| n,\alpha\rangle$ one obtains
\begin{equation}
\hat k_b^{(N)}|n,\alpha\rangle = \sum_{m(\neq
  n)}\frac{(-1)^{m-n}}{i(m-n)a}|m,\alpha\rangle. 
\end{equation}
As the $|\langle m,\alpha|\hat k_b^{(N)}|n,\alpha\rangle|$ decay as $1/|m-n|$ for
$|m-n|\to \infty$ , the states $\hat k_b|n,\alpha\rangle$ lie in the
Hilbert space ${\cal H}_N$ spanned by the Wannier states of the $N$
bands. If one would take Eq. (40) as the {\it definition} of the
operator $\hat k_b$ it is straightforward to show using Eqs. (33) and
(40) that 
\begin{eqnarray}
\hat k_b^{(N)}|k,\alpha\rangle 
& = & 
\left(2 \sum^\infty_{l=1} \frac{(-1)^{l+1}}{la} \sin
  kla\right)|k,\alpha\rangle\nonumber\\[0.3cm]
& \equiv &
f_{st}(k)|k,\alpha\rangle, 
\end{eqnarray}
where $f_{st}(k)$ is the periodic saw tooth curve which is given by
$f_{st}(k) = k$ in the first Brillouin zone.
The use of Eq. (40) also allows an alternative
derivation of the restricted Hilbert space commutation relation
corresponding to Eq. (27). For that purpose we decompose the position
operator using Eq. (30) as $\hat x^{(N)} \equiv \hat x_d^{(N)} + \hat
x_\Omega^{(N)}$ with 
\begin{equation}
\hat x_d^{(N)} = \sum^N_{i=1} \sum_m |m,\alpha_i\rangle am \langle m,\alpha_i|\;.
\end{equation}
This part of the position operator, which is diagonal in the Wannier
representation, leads to the nonvanishing commutator with $\hat
k_b^{(N)}$. As $\hat x_d^{(N)}|m,\alpha_i\rangle = a m |m,\alpha_i\rangle$ the
states $\hat x_d^{(N)}|m,\alpha_i\rangle$ as well as the states $\hat
k_b^{(N)}\hat x_d^{(N)}|m,\alpha_i\rangle$ are normalizable and therefore
elements of ${\cal H}_N$. If on the other hand $\hat k_b^{(N)}$ is
applied first, one obtains
\begin{equation}
\hat x_d^{(N)}\hat k_b^{(N)}|n,\alpha_i\rangle = \sum_{m (\neq n)}
\frac{(-1)^{m-n}m}{i(m-n)}|m,\alpha_i\rangle.
\end{equation}
These states are {\it not} normalizable and therefore (like the Bloch
states) strictly speaking {\it not} elements of the Hilbert space
${\cal H}_N$. If one nevertheless formally calculates the commutator
of $\hat x_d^{(N)}$ and $\hat k_b^{(N)}$ it is given by
\begin{eqnarray}
\left[ \hat x_d^{(N)}, \hat k_b^{(N)}\right]|n,\alpha_i \rangle
& = & 
-i \sum_{m(\neq n)}(-1)^{m-n}|m,\alpha_i\rangle \\
& = &
i\left(|n,\alpha_i\rangle - (-1)^n \sum_m
  (-1)^m|m,\alpha_i\rangle\right).\nonumber
\end{eqnarray}
The first form shows e.g. that the diagonal elements in the Wannier
representation of the commutator {\it vanish}, i.e. it {\it cannot} be
proportional to the unit operator. If Eq. (44) is multiplied by
$\langle n,\alpha_i|$ from the right and summations over $n$ and $i$ are
carried out, the expression for the commutator reads 
\begin{equation}
\left[\hat x_d^{(N)}, \hat k_b^{(N)}\right] 
= i\left(\hat 1_N - \frac{2\pi}{a} \sum_{i=1}^N |k = \frac{\pi}{a}, \alpha_i
\rangle \langle k = \frac{\pi}{a}, \alpha_i|\right),
\end{equation}
which is the analog of Eq. (27) for the case of the restricted
 Hilbert space. As in
section II it could have been derived also by using the derivative of
Eq. (36) at $t=0$ and using Eq. (31) to see that $\hat x_\Omega^{(N)}$
commutes with $\hat k_b^{(N)}$. In the next section we discuss how the
above commutation relation occurs in a completely different
context. There the analog of the Wannier states is the starting point
of the description.

\section{Bosonization and Klein operators}
In this section we first present a very short introduction to the
ideas behind bosonization, which is a very successful technique for
treating interacting fermions in one dimension \cite{Ha,GNT,DS,KS}. Consider a
system of fermions with one-particle basis states $|l\rangle$ and the
corresponding annihilation operators $c_l$, where $\l \in \mathbf Z$
is a quantum number running from $-M_0$ to $\infty$. For spinless
particles in a $1d$ box with hard walls \cite{SM} $M_0 = -1$. In the
treatment for interacting fermions Luttinger \cite{L} introduced the
purely technical device to add ``unphysical'' one particle states with
negative $l$ and considered the limit $M_0 \to \infty$. In order to
avoid mathematical subtleties it is more transparent to introduce the
concept of bosonization for {\it finite} $M_0$ and to perform the limit
$M_0 \to \infty$ later \cite{KS}. The state where the lowest one
particle states from  $-M_0$ to $N$ are occupied is denoted by
$|\{0\},N\rangle$, where the symbol $\{0\}$ is introduced, because instead of
using fermionic occupation numbers it is possible to work with
$N+M_0+1$-particle basis states 
\begin{equation}
|\{m_l\},N\rangle = \prod_{l>0}\frac{(b_l^\dagger)^{m_l}}{\sqrt{m_l!}}|\{0\},N\rangle,
\end{equation}
where the operators $b^\dagger_l$ with $l\ge 1$ are defined as
\begin{equation}
b^\dagger_l \equiv \frac{1}{\sqrt{l}}
\sum^\infty_{n=-M_0}c^\dagger_{n+l} c_n.
\end{equation}
In the limit $M_0 \to \infty$ these operators obey Bose commutation
relations $[b^\dagger_l, b^\dagger_{l'}]= 0$ and
\begin{equation}
[b_l,b^\dagger_{l'}] = \delta_{ll'}\hat 1,
\end{equation}
while for finite $M_0$ Eqs. (46) and (48) only hold for {\it low lying
  excited states} \cite{To,KS}. If the fermions have spin, all
operators have additional spin labels. For the point about ``Klein
factors'' we want to make, it is sufficient to work with a {\it
  single} branch of fermions as in Eqs. (46)-(48). The Klein factors
appear in the attempt to express the annihilation operator $c_l$ in
terms of the Bose operators defined in Eq. (47) \cite{Ha,DS,KS}. It
turns out to be easier to work with the auxiliary field operator
$\tilde \psi(v)$ \cite{KS} which in the limit $M_0 \to \infty$ is
defined as 
\begin{equation}
\tilde \psi(v) = \sum^\infty_{l = - \infty}e^{ilv} c_l\; .
\end{equation}
Its commutation relations with the Bose operators are
\begin{eqnarray}
\left[b_m,\tilde \psi(v)\right] 
&=&
 -\frac{1}{\sqrt{m}}e^{-imv} \tilde \psi(v),\;\nonumber\\
\left[b^\dagger_m,\tilde \psi(v)\right]
&=&
-\frac{1}{\sqrt{m}}e^{imv}\tilde\psi(v).
\end{eqnarray}
The essential observation for the ``bosonization'' of $\tilde \psi$ is
the fact that exponentials of Bose operators obey analogous commutation
relations
\begin{eqnarray}
\left[b_m, e^{-\sum^{\infty}_{n=1} \frac{e^{-inv}}{\sqrt{n}}b^\dagger_n}\right]
&=&
- \frac{1}{\sqrt m}\;e^{-imv}e^{-\sum^\infty_{n=1}\frac{e^{-inv}}{\sqrt
      n}b^\dagger_n}\\ \nonumber
\left[b_m^\dagger,e^{\sum^{\infty}_{n=1}\frac{e^{inv}}{\sqrt{n}}b_n}\right]
&=&
- \frac{1}{\sqrt m}\;e^{imv}\;e^{\sum^\infty_{n=1}\frac{e^{inv}}{\sqrt
      n}b_n}.
\end{eqnarray}
If one therefore defines the ``Klein operator'' $\hat O(v)$ as follows
\begin{equation}
\tilde \psi(v) \equiv \hat O(v) e^{-\sum^{\infty}_{n=1}
\frac{e^{-inv}}{\sqrt{n}}b^\dagger_n}
e^{\sum^\infty_{n=1}\frac{e^{inv}}{\sqrt n}b_n},
\end{equation}
Eqs. (50) and (51) imply that $\hat O (v)${\it commutes} with the
$b_n$ and $b^\dagger_n$. As the exponentials in Eq. (52) conserve the
particle number, the Klein operator $\hat O(v)$ has to lower the
particle number by one as the field operator on the lhs of the
equation. In the following an explicit construction of $\hat O (v)$ is
presented, which considerably simplifies earlier approaches
\cite{Ha}. The first step is to show that the
 state $\hat O(v)|\{0\},N\rangle $ has no overlap to excited
states. Using 
\begin{eqnarray}
\langle \{m_l\},N-1|\hat O(v)|\{0\},N\rangle
 =  \qquad \qquad \qquad \\
\prod_{l}\frac{1}{\sqrt{m_l!}}
\langle \{0\},
N-1|\left(b_l\right)^{m_l}\hat O(v)|\{0\},N\rangle\nonumber
\end{eqnarray}
and the fact that $\hat O(v)$ commutes with the $b_l$, the rhs of
Eq. (53) is seen to vanish unless all $m_l$ are zero,
as $b_l|\{0\},N\rangle=0$. This implies
\begin{equation}
\hat O(v)|\{0\},N\rangle = c_N(v)|\{0\},N-1\rangle,
\end{equation}
where $c_N(v)$ is a $c$-number. In order to determine $c_N(v)$ we
calculate $\langle \{0\}, N-1|\tilde \psi(v)|\{0\}N\rangle$
 using Eq. (49) and compare
with the result when Eqs. (52) and (54) are used. This yields $c_N(v) =
e^{iNv}$. If Eq. (54) is multiplied by
$\prod_{l} (b_l^\dagger)^m/\sqrt{m_l!}$ from the left, the fact that 
$\hat O(v)$ commutes with the $b^\dagger_l$ completely determines
$\hat O (v)$
\begin{equation}
\hat O(v)|\{m_l\},N\rangle = e^{iNv}|\{m_l\},N-1\rangle\,.
\end{equation}
Using the completeness of the states $|\{m_l\},N\rangle$ one can write
Eq. (55) in the form
\begin{equation}
\hat O(v) = \hat U e^{i\hat{\cal N}v}
\end{equation}
where $\hat{\cal N}$ is the particle number operator relative to the
Dirac see
\begin{equation}
\hat{\cal N} = \sum_{\{m_l\}} \sum_N |\{m_l\},N\rangle N\langle \{m_l\},N|
\end{equation}
and the particle number changing part $\hat U$ which is {\it
  independent} of $v$ is given by
\begin{equation}
\hat U = \sum_{\{m_l\}}|\{m_l\}, N-1\rangle\langle \{m_l\},N|
\end{equation}
If one identifies the particle number $N$ with the site index and the
$\{m_l\}$ with the $\alpha_i$ in Eq. (32), the operator $\hat U$ has the
form of a translation operator $T_{-a}$, which shifts the system to
the left by one lattice site. Therefore the eigenstates of $\hat U$
are the corresponding ``Bloch states''
\begin{equation}
|\{m_l\},k\rangle \equiv \frac{1}{\sqrt{2\pi}} \sum^\infty_{N= -\infty}
e^{ikN} |\{m_l\},N\rangle\; .
\end{equation}
Now one can again write $\hat U = e^{i\hat k}$ with
\begin{equation}
\hat k \equiv \sum_{\{m_l\}}\int^\pi_{-\pi}|\{m_l\},k\rangle k\langle \{m_l\},k|dk.
\end{equation}
As $\hat{\cal N}$ corresponds to $\hat x_d^{(N)}$ in section III, the
same steps as in calculation of $\left[\hat x_d^{(N)}, \hat
  k_b^{(N)}\right]$ lead to \cite{KS}
\begin{equation}
[\hat{\cal N}, \hat k] = i\left(\hat 1 - 2\pi \sum_{\{m_l\}}|\{m_l\},
  k=\pi\rangle\langle \{m_l\}, k = \pi| \right)\;,
\end{equation}
which {\it differs} from what is usually assumed \cite{Ha,GNT} by the
second term on the rhs \cite {HaII}.

\section{Implications}
\noindent Here we want to discuss if the subtleties in the definitions of the
$\hat k$ operators and their non-canonical commutation relations
have practical implications. We begin 
with the case of a Bloch electron in a homogenous field.
Let us first summarize the main results presented in sections II and
III. The well known fact that
an initial state in the form of a linear combination of Bloch states
with different band indices but the {\it same} value of $k$ remains
in such a form with a sharp value of $k$, was expressed as 
$\hat P_k(t) = \hat P_{k-Ft/\hbar }(0)$
(Eq. (24)). But in contrast to common
belief this does {\it not} lead to Newton's law in the form of Eq. (1)
if $k(t)$ is interpreted as the expectation value of the linear
{\it operator} $\hat k_b(t)$ with $\hat k_b$ defined in Eq. (34).
The property usually not correctly taken care of is the {\it periodicity}
in the reciprocal lattice
of the operator $\hat P_k$ (Eqs. (22) or (34)) .
An obvious way to avoid contradictions is {\it not} to work with  $\hat k_b$
at all, but take Eq. (24) as the basic result and use the
translation operator $T_a$ {\it only}, 
e.g. in the form of
Eq. (32) {\it without} introducing a $\hat k$ operator. 
We briefly show that this is a useful way to proceed in the single band case.
The Hamiltonian $\hat H_0$ can be expressed in terms of $\hat T_a^{(1)}$
\begin{equation}
\hat H_0=\sum_{n>0}\left(\epsilon_{n,0}(\hat T^{(1)}_a)^n +h.c.\right),
\end{equation}
where the $\epsilon_{n,0}$ are the hopping matrix elements. In the
following we drop the superscript $(1)$ at the single band translation 
operator $ \hat T^{(1)}_a $         . Its representation in Eq. (32)
 easily leads to the
commutation relation $[\hat x, \hat T_a^n]=na\hat T_a^n$, where
$\hat x \equiv \hat x_d^{(1)}$. Therefore the Heisenberg equation
of motion
for $\hat x(t)$ reads
\begin{equation}
i\hbar \dot {\hat x}(t)=a\sum_{n>0}\left( n\epsilon_{n,0}\hat T_a^n(t)
-h.c.\right).
\end{equation}
With Eq. (4) its solution can be found by integrating from $0$ to $t$
\begin{equation}
\hat x(t)-\hat x = \sum_{n>0}\left [     \epsilon_{n,0}
\hat T_a^n \left (e^{-inaFt/\hbar} -1\right)/F  + h.c.\right]  .       
\end{equation}
If one takes an arbitrary wave packet as an initial state the
expectation value of $\delta\hat x(t)\equiv \hat x(t) -\hat x$ shows
{\it periodic} behaviour with time $T=2\pi\hbar/aF$, the well known
{\it Bloch oscillations}. For the case of nearest neighbour hopping
 the expectation value of $\hat T_a$ determines the amplitude of the
oscillation. If the wave packet is well localized in $k$-space the
amplitude is given by $B/2F$, where $B$ is the band width, while 
a Wannier-Stark eigenstate $|E^{\mu}\rangle$ or a Wannier state $|m \rangle$
as a initial state lead to {\it zero} amplitude of the Bloch oscillation. 
For the former all expectations are time independent, while the
Wannier state behaves similar to a squeezed state with a periodic
modulation of $\langle (\delta \hat x(t))^2\rangle$. Note that for
this description of the $N=1$ case {\it no} $\hat k$-operator was
needed. For $N>1$ the Bloch oscillations generically are only
{\it quasiperiodic} and numerical methods have to be used to determine
their form\cite{H,Z}. Again one can completely avoid to introduce $\hat k_b$.  
   
\noindent In the case of the Klein operators there exist examples 
where the analogous idea to completely avoid phase operators was
advocated \cite{DS}. But the usual procedure is to introduce the
phase operators and to {\it neglect} the last term on the rhs of
Eq. (61), i.e. to assume canonical commutation relations for the
particle number and the phase operator \cite{Ha}
in order to be able to apply powerful
field theory techniques\cite{GNT}.
 It is known that this procedure leads to the correct
anticommutation relations for the field operators Eq. (49),
\cite {Ha} but it should be
obvious that it can also lead to errors. As an example we 
discuss dynamical properties of a system described by a Hamiltonian 
\begin{equation}
\hat H=\hat H_B+f( \hat {\cal N}),
\end {equation}
where $\hat H_B$ is a function of the boson operators $b_n$ and
$b_n^{\dagger}$ and $f$ is an arbitrary function specified later. 
Note that the two terms in the Hamiltonian commute.
A typical example for this form is the Tomonaga-Luttinger model \cite
{Ha,DS,KS}.
As a first exercise we calculate $\hat U(t)=e^{i\hat Ht}\hat U
e^{-i\hat Ht} $
(putting $\hbar =1$)  
in two different ways.
This operator is an important ingredient for the calculation of 
fermionic correlation functions of the model. 
 Using the definition
Eq. (58) the exact evaluation is simple as the $    |\{m_l\}, N\rangle\  $
are eigenstates of $    \hat {\cal N}   $, and $\hat H_B$
commutes with $\hat U$. This yields
\begin{equation}
\hat U(t)=\hat U e^{-i[f( \hat {\cal N}) - f( \hat {\cal N}-1)]t }.
\end{equation}
Next we calculate $\hat U(t)=e^{i \hat k(t)}$ by calculating 
$\hat k(t)$ using the Heisenberg equation of motion and the
{\it wrong} commutation relation $ [\hat{\cal N}, \hat k] = i\hat 1
$, which gives $[\hat k,H]=-if'(\hat {\cal N})$. As
$\hat {\cal N}$ is a constant of motion this yields 
$\hat k(t)_{CR}=\hat k -if'(\hat {\cal N})t$, where the index $CR$ is
a reminder that we have used the wrong commutation
relation. This shows that the expectation value of $\hat k(t)_{CR} $
increases linearly with time in eigenstates $ |E^B_i,N\rangle $
of $\hat H$, while the expectation value of the exact $\hat k(t)$ is
time
{\it independent}. 
If we put
the incorrect result $\hat k(t)_{CR} $ into $e^{i\hat k(t)}$ we obtain
\begin{eqnarray}
\hat U(t)_{CR}
& = & e^{i[\hat k -f'(\hat {\cal N})t]} \nonumber \\
& = & e^{i\hat k} \hat T_{\alpha} \left ( e^{-it\int_0^1 f'(\hat
    {\cal N} (\alpha))d\alpha }\right ),
\end{eqnarray}
where  $ \hat {\cal N} (\alpha)\equiv e^{i\alpha \hat k} \hat
{\cal N} e^{-i\alpha \hat k}$ and $\hat T_{\alpha}$ is the ``time''($\alpha$)
ordering symbol. Again using the wrong commutation relation yields
$d\hat {\cal N}(\alpha)/d\alpha=-i\hat 1$, i.e. $ \hat {\cal
  N}(\alpha)=\hat {\cal N}-\alpha \hat 1$. As the $ \tilde {\cal
  N}(\alpha)   $ for different arguments {\it commute}, the ordering
symbol $\hat T_{\alpha}$ in Eq. (67) can be deleted and performing 
 the integration shows that $\hat U(t)_{CR}$ {\it agrees} with the exact
 result for  $\hat U(t)$ in Eq. (66).
 For the Tomonaga-Luttinger  
model $f$ is a second order polynomial \cite {Ha,DS,KS}.
Then $f'(\hat {\cal N})$ is linear in $ \hat {\cal N}$ and 
the Baker-Hausdorff relation can be used for an alternative proof
of $\hat U(t)=\hat U(t)_{CR} $ for Hamiltonians of the type 
presented in Eq. (65).
The fact that the incorrect procedure gives the {\it exact}
result for $\hat U(t)$, shows that
for the calculation of fermionic correlation functions  
one is lucky 
for this ``fixed point model'' of interacting fermions in one
dimension.
If backscattering or impurity scattering are included the Hamiltonian 
has additional terms which are not of the simple form
of Eq. (65). If one wants e.g. to evaluate the partition function for
a Hamiltonian $\hat H +\lambda \hat H'$ the expression in perturbation
theory to all orders in $\lambda$ involves correlation functions for
the system described by $\hat H$ defined in Eq. (65). The arguments
just presented show that  
in this case the use of the wrong commutation relation
does not lead to incorrect results.  

\noindent Nevertheless the example presented shows that the widely accepted
use of the wrong commutation relation between the phase and the
particle number operator
requires proper care.
\section{Acknowledgements}
The author would like to thank G. Hegerfeldt, P. Kopietz, V. Meden,
A. Nersesyan and W. Zwerger for useful discussions.
 
\begin{appendix}
\section{}
\noindent In this appendix we present the simple proof of Eq. (22) which states
that the operators $\hat P_k$ originally defined in terms of plane wave
states also have a simple representation in terms of Bloch states of
an arbitrary periodic potential. In the proof we use the well-known
identity 
\begin{equation}
\sum^\infty_{m = -\infty} e^{imu} = 2 \pi
\sum^\infty_{l=-\infty}\delta (u + 2\pi l)\equiv 2\pi \delta_{2\pi}(u),
\end{equation}
where the index ``$2\pi$'' indicates the $2\pi$-periodicity of the
argument of the delta function. 
We use this relation to show how $\hat P_k$ acts in the position
representation
\begin{eqnarray}
\langle x|\hat P_k|\psi \rangle
& = &
\int_{-\infty}^{\infty}\frac {1}{2\pi}\sum_{n=-\infty}^{\infty}
e^{i\frac {2\pi}{a} n(x-x')}
 e^{ik(x-x')}
  \psi(x') dx'\nonumber\\
& = &
\frac {a}{2\pi} \sum_{m=-\infty}^{\infty}e^{ikam}
  \psi(x-am).
\end{eqnarray}
For a Bloch state $|k',\alpha \rangle$ Bloch's theorem 
$\langle x-am|k',\alpha \rangle
= e^{-ik'am}          \langle x  |k',\alpha \rangle 
$ in Eq. (A2) yields, again   using
Eq. (A1)
\begin{equation}
\langle x|\hat P_k|k',\alpha \rangle =
\delta_{2\pi /a}(k-k')\langle x|k',\alpha \rangle.
\end{equation}
If we define the operator $\hat{\tilde P}_k \equiv \sum_\beta
|k,\beta\rangle\langle k,\beta|$ the matrix elements $\langle x|\hat{\tilde
  P}_k|k',\alpha\rangle$ follow immediately as
\begin{equation}
\langle  x|\hat{\tilde P}_k|k',\alpha\rangle =
\delta_{2\pi/a}(k-k')\langle
 x|k',\alpha\rangle.
\end{equation}
As $\langle x|$ and $|k',\alpha\rangle     $ in Eqs. (A3) and (A4)
 are {\it arbitrary}
this proves $\hat P_k = \hat{\tilde P}_k$, i.e. Eq. (22).
\end{appendix}


\begin{thebibliography}{*}
\bibitem{Ki} C. Kittel, {\it Introduction to Solid State Physics} (John
  Wiley, New York, 1967) 
\bibitem{AM} N. Ashcroft and D. Mermin {\it Solid State
    Physics} (Holton \& Rice, New York, 1976) 
\bibitem{QKi} C. Kittel, {\it Quantum Theory of Solids} (John Wiley, New
  York, 1963) 
\bibitem{Ca} J. Callaway, {\it Quantum Theory of the Solid State, Part
    B} (Academic, New York, 1974)
\bibitem{Av} J.E. Avron, Ann. Phys. (N.Y.){\bf 143}, 33 (1982)
\bibitem{Ne} G. Nenciu, Rev. Mod. Phys. {\bf 63}, 91 (1991)
\bibitem{Wa} G.H. Wannier, Phys. Rev. {\bf 117}, 1366 (1960)
\bibitem{Kri} J.B. Krieger and G.J. Iafrate, Phys. Rev. {\bf B33}, 5494
  (1986)
\bibitem{J} D. Judge, Phys. Lett. {\bf 5}, 189 (196)
\bibitem{K} K. Kraus, Z. Phys. {\bf 188}, 374 (1965) and unpublished
  notes
\bibitem{Kr} H. Kroemer, Am. J. Phys. {\bf 54}, 177 (1986)
\bibitem{Ha} F.D.M. Haldane, J. Phys. {\bf C14}, 2585 (1981)
\bibitem{GNT} A.O. Gogolin, A.A. Nersesyan and A.
         M. Tsvelik, {\it Bosonization and strongly correlated
         systems} (Cambridge University Press, 1998)
\bibitem{DS} J. von Delft and H. Schoeller, Ann. Phys. (Leipzig) {\bf
    7}, 225 (1998)
\bibitem{KS} K. Sch\"onhammer, ``Interacting Fermions in one dimension:
  The Tomonaga-Luttinger model'', cond-mat/ 9710330. This pedagogical
  review was
  intended as part 2 of reference \cite{SM}, but was considered too
  technical for the general reader by the editors of Am. J. Phys.
\bibitem{SM} K. Sch\"onhammer and V. Meden, Am. J. Phys. {\bf 64}, 1168
  (1996)
\bibitem{CaII} Were the ``usual'' argument fails can be most clearly
  seen in Ref. 4, where a discussion of the general solution $|\psi
  (t)\rangle$ of the time dependent Schr\"odinger equation in the
homogenous external field is given. The author fails to take into
account that $\sum_{\alpha}|\langle k,\alpha|\psi (t)\rangle|^2$
is a {\it periodic} funtion of $k$ in his calculation
of $\langle k\rangle$.  
\bibitem{Beispiel} As a simple example we take
$\hat {\tilde k}_b =\hat k_b +\int_{k_1}^{k_2}\hat P_k dk$. Using
$[\hat x, \hat P_k]=-i \partial \hat P_k/\partial k$ this yields
$[\hat x ,\hat {\tilde k}_b]=i \left [\hat 1 -(2\pi/a)(
\hat P_{\pi/a}+\hat P_{k_2}-\hat P_{k_1})\right ]$. For a finite
number of additional integrals in the definition of  $\hat {\tilde k}_b $
one obtains a corresponding additional sum in the commutator.
\bibitem{FBF} H. Fukuyama, R.A. Bari and H.C. Fogedby, Phys. Rev. {\bf
    8}, 5579 (1973)
\bibitem{H} D.W. Hone and X.G. Zhao, Phys. Rev. {\bf B53} 4834 (1996)
\bibitem{Z} X.G. Zhao, G.A. Georgakis and Q. Nin, Phys. Rev. {\bf
    B54}, R5235 (1996)
\bibitem{L} J.M. Luttinger, J. Math. Phys. {\bf 4}, 1154 (1963)
\bibitem{To} S. Tomonaga, Prog. Theor. Phys. {\bf 5}, 544 (1950)
\bibitem{HaII} In his seminal paper \cite {Ha} where 
he first discussed the universality
of the ``Luttinger liquid'' concept, Haldane also
pointed out the necessity of the 
proper inclusion of what are now called ``Klein factors''
in the bosonization of the electronic field operators.  
His bosonization formula (Eq. (3.37) his absolutely correct, but
the first commutator relation in his Eq. (3.24) is {\it not}. 
How one can nevertheless try to avoid errors is discussed in section V.

          
\end{thebibliography}
\end{document}